\documentclass[aps,pre,twocolumn,superscriptaddress,showpacs,floatfix]{revtex4-1}
\usepackage{dcolumn}
\usepackage{amsmath}
\usepackage{amssymb}
\usepackage{amsthm}
\usepackage{graphicx}
\usepackage{bm}
\usepackage[T1]{fontenc}
\usepackage{color}
\usepackage{url}
\usepackage[bf]{subfigure}
\usepackage{rotating}
\usepackage{scalefnt}
\usepackage{multirow}

\usepackage{scalefnt}
\usepackage{times}
\usepackage{mathtools}
\usepackage{bbm}

\usepackage{enumerate}

\usepackage{tikz}
\usetikzlibrary{topaths,calc}

\theoremstyle{definition}

\newcommand{\A}{\mathbf{A}}
\newcommand{\W}{\mathbf{W}}

\newcommand{\D}{\mathbf{D}}

\newcommand{\F}{\mathbf{F}}
\newcommand{\G}{\mathbf{G}}

\newcommand{\M}{\mathbf{M}}

\newcommand{\Lap}{\mathbf{L}}

\newcommand{\x}{\mathbf{x}}

\DeclareMathOperator{\I}{\mathbf{I}}

\newcommand{\vertii}[1]{{\left\vert\kern-0.25ex\left\vert #1 \right\vert\kern-0.25ex\right\vert}}

\newcommand{\slfrac}[2]{\left.#1\middle/#2\right.}

\begin{document}

\title{Phase transitions and stability of dynamical processes on hypergraphs}

\author{Guilherme Ferraz de Arruda}
\thanks{These two authors contributed equally}
\affiliation{ISI Foundation, Via Chisola 5, 10126 Torino, Italy}

\author{Michele Tizzani}
\thanks{These two authors contributed equally}
\affiliation{ISI Foundation, Via Chisola 5, 10126 Torino, Italy}

\author{Yamir Moreno}
\affiliation{Institute for Biocomputation and Physics of Complex Systems (BIFI), University of Zaragoza, 50018 Zaragoza, Spain}
\affiliation{Department of Theoretical Physics, University of Zaragoza, 50018 Zaragoza, Spain}
\affiliation{ISI Foundation, Via Chisola 5, 10126 Torino, Italy}

\begin{abstract}
Hypergraphs naturally represent higher-order interactions, which persistently appear from social interactions to neural networks and other natural systems. Although their importance is well recognized, a theoretical framework to describe general dynamical processes on hypergraphs is not available yet. In this paper, we bridge this gap and derive expressions for the stability of dynamical systems defined on an arbitrary hypergraph. The framework allows us to reveal that, near the fixed point, the relevant structure is the graph-projection of the hypergraph and that it is possible to identify the role of each structural order for a given process. We also analytically solve two dynamics of general interest, namely, social contagion and diffusion processes, and show that the stability conditions can be decoupled in structural and dynamical components. Our results show that in social contagion process, only pairwise interactions play a role in the stability of the absorbing state, while for the diffusion dynamics, the order of the interactions plays a differential role. Ours is the first attempt to provide a general framework for further exploration of dynamical processes on hypergraphs.
\end{abstract}

\maketitle

Network science has been successful in describing the structure and dynamics of complex systems in many fields of science. Nevertheless, in the vast majority of studies, the mathematical and computational descriptions of these networks are limited to pairwise interactions. In most cases, this order of interaction is a good approximation, describing real phenomena~\cite{Boccaletti2008, Newman010:book, Satorras2015, cimini2019}. Recently, it has become increasingly evident that simplifying higher-order interactions as a set of pairwise ones can be misleading in many situations~\cite{Lambiotte2019, Chodrow2019}, as it is the case, for instance, when comparing triangles and clustering~\cite{Chodrow2019}. An attempt to formally solve this problem is represented by the introduction of the simplicial complexes approach~\cite{Petri2018, Iacopo2019}. Although very powerful,
as a topological space, simplicial complexes require mutual inclusion for the interactions~\cite{armstrong2013basic}, which is too restrictive for general purposes. To face this problem, one needs to resort to the use of hypergraphs, which by relaxing the assumption of mutual inclusion, allow the representation of a broader range of systems.

Admittedly, the attention to higher-order systems has proliferated lately~
\cite{Estrada2006, Ghoshal2009, Bodo2016, Banerjee2017, benson2018simplicial, Banerjee2019, Lambiotte2019, Iacopo2019, Chodrow2019, Ouvrard2018, Arruda2019, 
Unai2020, Carletti2019}, with an increasing focus on hypergraphs in fields such as mathematics~\cite{Bodo2016, qi2017, Banerjee2017, Banerjee2019, Chodrow2019, Ouvrard2018, ouvrard2020}, physics~\cite{Estrada2006, Ghoshal2009, Arruda2019, Unai2020, Carletti2019}, and computer science \cite{Karypis1999, Tian2009, Voloshin2013, Bretto2013, Valdivia2019, Payne2019, Jiang2019}. Despite this interest, a general theory of dynamical processes on higher-order structures is still largely missing. To fill this important gap, here we build a mathematical framework that allows performing a linear stability analysis for general processes on arbitrary hypergraphs. This approach highlights the importance of the graph-projection $-$an underlying weighted graph representation of a hypergraph$-$ for the dynamics. The proposed methodology makes it possible to decompose a hypergraph in uniform structures, hence enabling the characterization of their role in the system's dynamics. Finally, to show the usefulness of our approach, we analytically study and recover some results reported for social contagion~\cite{Iacopo2019, Arruda2019}, and diffusion processes, also providing new key insights into these paradigmatic dynamics. The framework discussed in this paper could be used to explore different processes when pairwise interactions are an oversimplification of the system, which has the potential to bring more insights into the understanding of higher order dynamics of interacting systems.

\begin{figure}[t]
\includegraphics[width=\linewidth]{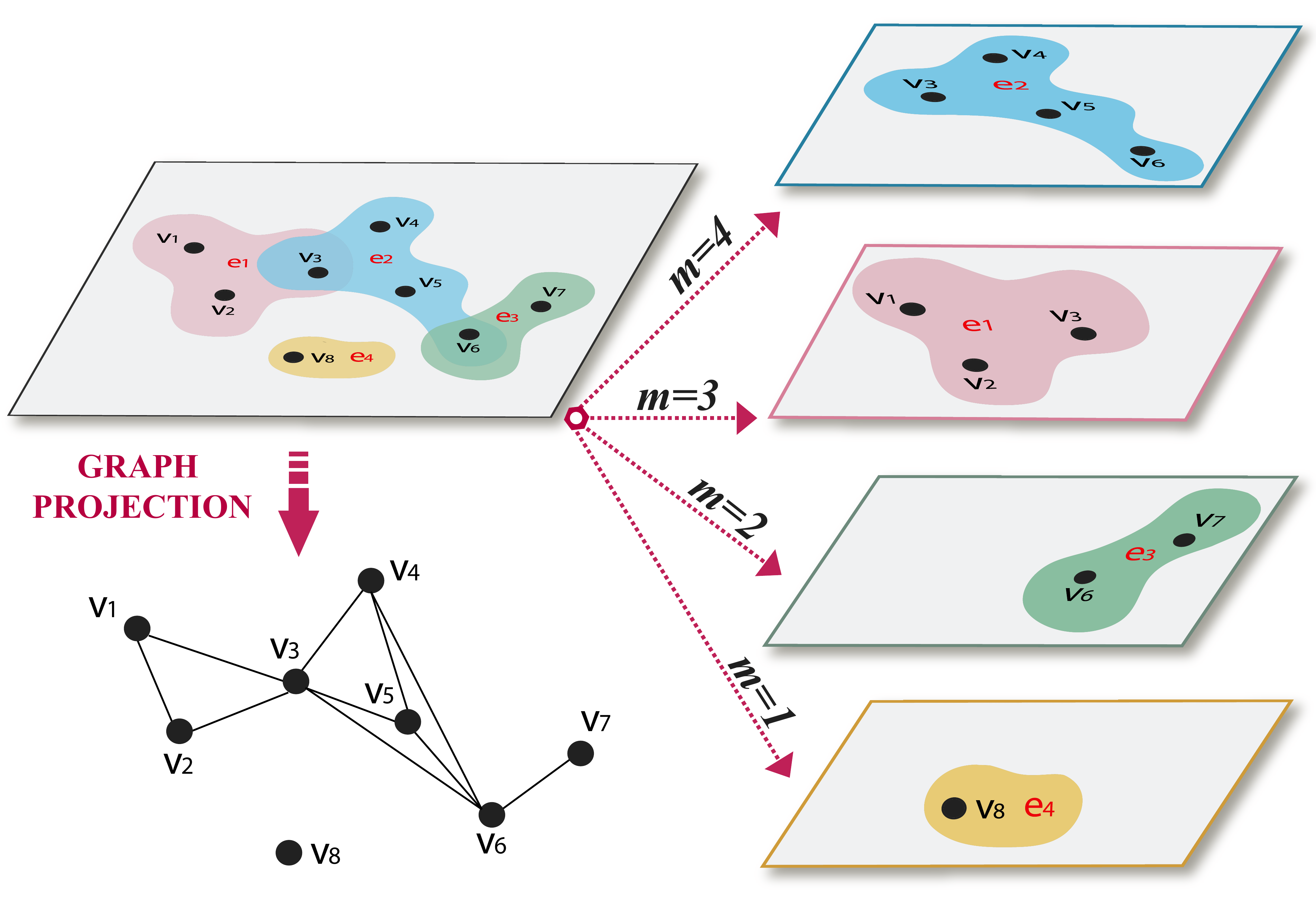}
\caption{Graphical representation of a hypergraph, its structural decomposition (right multilayer panel) and projection (bottom graph). Mathematically, $\mathcal{V} = \{v_1, v_2, v_3, v_4, v_5, v_6, v_7, v_8 \}$, $\mathcal{E} = \{e_1, e_2, e_3, e_4 \}$, where the hyperedges are $e_1 = \{v_1, v_2, v_3 \}$, $e_2 = \{v_3, v_4, v_5, v_6 \}$, $e_3 = \{v_6, v_7 \}$ and $e_4 = \{v_8 \}$. We show how the original hypergraph can be represented by decomposing it into $m$-uniform hypergraphs or projected $-$where hyperedges are simplified as $|e_j|$-cliques.}
\label{fig:hypergraph}
\end{figure}

\textit{Hypergraph structure.} 
A hypergraph, $\mathcal{H} = \{ \mathcal{V}, \mathcal{E}\}$, is defined as a set of nodes, $\mathcal{V} = \{ v_i \}$, with $N = |\mathcal{V}|$ the number of nodes and a set of hyperedges $\mathcal{E} = \{ e_j \}$, where $e_j$ is a subset of $\mathcal{V}$ with arbitrary cardinality $|e_j|$. Note that if $\max \left( |e_j| \right) = 2$ we recover a graph, whereas one has a simplicial complex if for each hyperedge with $|e_j| > 2$, its subsets are also contained in $\mathcal{E}$. The adjacency matrix~\cite{Banerjee2019} can be defined as 
\begin{equation} \label{eq:adjacency}
    \A_{ij} = \sum_{\substack{e_j \in \mathcal{E} \\ i,j \in e_j}} \frac{1}{|e_j| - 1},
\end{equation}
which can be interpreted as a weighted projected graph. An example of a hypergraph and its graph projection is shown in Fig.~\ref{fig:hypergraph}. The weighted counterpart of the adjacency matrix is defined as
\begin{equation} \label{eq:w_adjacency_m}
  \W_{ij} = \sum_{\substack{e_j \in \mathcal{E} \\ i,j \in e_j}} \frac{w(|e_j|)}{|e_j| - 1},
\end{equation}
where $w(|e_j|)$ is a function of the cardinality of the hyperedge, $e_j$, weighting differently the contribution of each hyperedge. For instance, if we consider $w(|e_j|) = |e_j| - 1$, we recover the adjacency matrix used in~\cite{Carletti2019}. Furthermore, the Laplacian matrix is defined as $\Lap = \D - \A$, where $\D = \text{diag} (k_i)$ and the degree is given as $k_i = \sum_{j=1}^N \A_{ij}$. Note that, in order to define the Laplacian matrix on the weighted case we also need to redefine $\D$ accordingly.

We can also decompose the hypergraph into $m$-uniform hypergraphs, as motivated in~\cite{Ouvrard2018}. Formally, 
\begin{equation} \label{eq:adjacency_m}
    \A_{ij}^m = \sum_{\substack{e_j \in \mathcal{E}, |e_j| = m  \\ i,j \in e_j}} \frac{1}{|e_j| - 1},
\end{equation}
therefore $\A = \sum_{m} \A^m$ as showed in Fig.~\ref{fig:hypergraph}. Consequently, we can also define $\D^m$, $\Lap^m$, and $\W^m$.

\textit{Stability analysis.} 
A general dynamical process on the hypergraph can be written as
\begin{equation} \label{eq:dxi}
    \dfrac{d x_i}{dt} = f_i (x_i) + \sum_{e_j \in \mathcal{E}_i} g_j (x_{\{e_j \}}),
\end{equation}
where $x_i$ is the state of the node $v_i$, $f_i (x_i)$ is a $\mathbb{R} \rightarrow \mathbb{R}$ function that depends only on the state of node $v_i$ and $g_j (x_{\{e_j \}})$ is a $\mathbb{R}^{|e_j|} \rightarrow \mathbb{R}$ function that takes all the states of the nodes on the hyperedge $e_j$, here denoted as $x_{\{e_j \}}$, and compute its contribution to $x_i$. While $f_i (x_i)$ represents the internal dynamics of the node, $g_j (x_{\{e_j \}})$ is the external interaction, expressed as the hyperedge contribution.

We can perform a linear stability analysis for Eq.~\ref{eq:dxi} around a known fixed point, $x_i = x_i^*$, similarly to~\cite{Newman010:book}. The linearized equations are expressed as
\begin{equation} \label{eq:eps}
    \dfrac{d \epsilon_i}{dt} = \dfrac{d f_i (x_i)}{dx_i} \bigg|_{x_i = x_i^*} \epsilon_i + \sum_{e_j \in \mathcal{E}_i} \sum_{k \in e_j} \partial_{x_k} g_j (x_{\{e_j \}}) \bigg|_{\textbf{x} = \textbf{x}^*} \epsilon_k,
\end{equation}
where $\textbf{x}$ is a vector whose components are $x_i$. We can express Eq.~\ref{eq:eps} in its matrix form decomposing a general hypergraph into uniform hypergraphs as
\begin{equation} \label{eq:eps_mat}
    \dfrac{d \mathbf{\epsilon}}{dt} = \left( \F(\x^*) + \sum_{m=1}^{\max \left( |e_j| \right)} \G^m(\x^*) \right) \mathbf{\epsilon} = \M \mathbf{\epsilon},
\end{equation}
where
\begin{eqnarray}
    \F_{ii}(\x^*) &=&
    \begin{cases}
        \dfrac{d f_i (x_i)}{dx_i} \bigg|_{x_i = x_i^*}  & i = j \\
        0 & \text{otherwise}
    \end{cases} \label{eq:F} \\
    \G_{ik}^m(\x^*) &=&  \sum_{\substack{e_j \in \mathcal{E}_i \\ |e_j| = m}} \partial_{x_k} g_j (x_{\{e_j \}}) \bigg|_{\textbf{x} = \textbf{x}^*}. \label{eq:G}
\end{eqnarray} 
The fixed point $\x^*$ is stable if 
\begin{equation} \label{eq:general_condition}
    \Lambda_i(\M) < 0 \hspace{2mm} \forall i \in \{ 1, 2, ..., N\},
\end{equation}
where $\Lambda_i(\M)$ is the $i$-th eigenvalue of $\M$. The structures of cardinality $m$ do not contribute to the stability of the fixed point $\x^*$ if $\G_{ik}^m(\x^*) = 0$ $\forall i,k$. We stress that, although we have a high-order structure ($N^{\max(|e_j|)}$ in the tensorial representation~\cite{Banerjee2017}), near the fixed point, the dynamics is described only by a $N^2$ matrix.

\begin{figure*}[!t]
\includegraphics[width=0.8\linewidth]{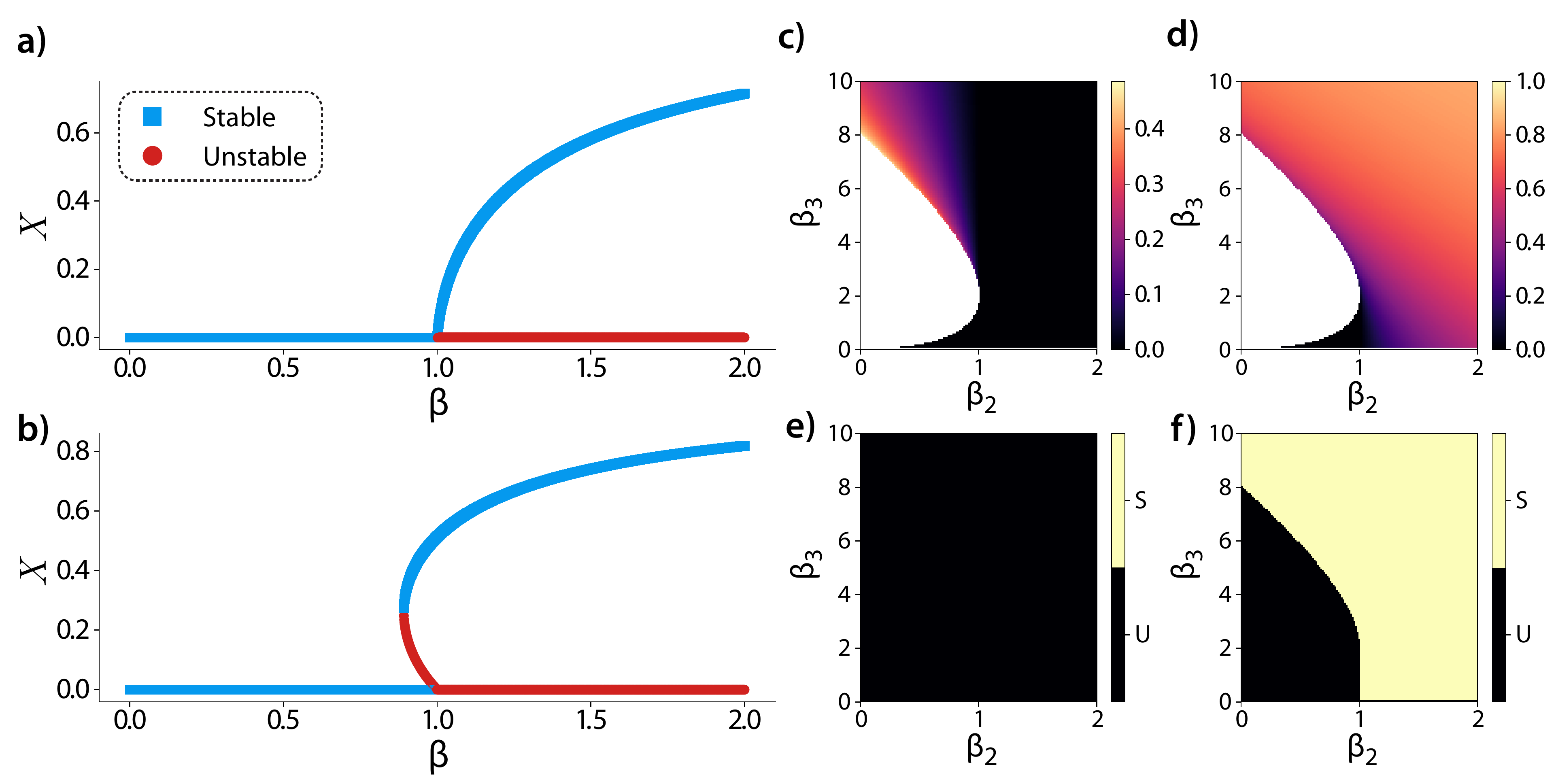}
\caption{Stability analysis of a social contagion dynamics on real-world (panels (a) and (b)) and synthetic hypergraphs (panels (c) through (f)). In panel (a), the substrate is a high school contact pattern, while in (b), a primary school contact pattern. Panels from (c) to (f) show results for the stability analysis of a simplicial complex. Panel (c) and (e) correspond, respectively, to the solution ${x^*}_2^-$ (Eq.~\ref{eq:social_cont_sol}) and its stability (Eq.~\ref{eq:social_cont_criteria}.), whereas (d) and (f) represent the solution ${x^*}_2^+$ (Eq.~\ref{eq:social_cont_sol}) and its stability (Eq.~\ref{eq:social_cont_criteria}), respectively.}
\label{fig:social_contagion}
\end{figure*}

Let us first consider a diffusion-like process. In this case, $g_j (x_{\{e_j \}})$ is separable and we can express the interaction function as
\begin{equation} \label{eq:eq_separable}
    g_j (x_{\{e_j \}}) = \frac{1}{|e_j| - 1} \sum_{\substack{k \in e_j \\ k\neq i}} \left( g_j (x_{i}) - g_j (x_{k})\right),
\end{equation}
where the weight $(|e_j| - 1)^{-1}$ is necessary due to conservation purposes. Assuming that $f_i(x_i)$ is the same for all nodes, the fixed point is symmetrical, i.e., $x_i^* = x_j^*$ for any pair $i$, $j$. In this case, Eq.~\ref{eq:F} reduces to  
\begin{equation} \label{eq:cond_1}
    \F = \alpha \I  \hspace{2mm} \text{and} \hspace{2mm} \G = \beta \A,
\end{equation}
where
\begin{equation} \label{eq:alpha_beta}
    \alpha = \dfrac{d f (x)}{dx} \bigg|_{\textbf{x} = \textbf{x}^*} \hspace{2mm} \text{and} \hspace{2mm} \beta =  \dfrac{d g_j (x)}{dx}\bigg|_{\textbf{x} = \textbf{x}^*}.
\end{equation}
Therefore, Eq.~\ref{eq:eps} is expressed as
\begin{equation} \label{eq:special_1}
 \dfrac{d \epsilon_i}{dt} = \left( \alpha + \sum_{e_j \in \mathcal{E}_i} \sum_{k \in e_j} \frac{\beta}{|e_j| - 1} \right) \epsilon_i + \sum_{e_j \in \mathcal{E}_i} \sum_{k \in e_j} \frac{\beta \epsilon_k}{|e_j| - 1},
\end{equation}
which in matricial form reads
\begin{equation}
 \dfrac{d \mathbf{\epsilon}}{dt} = \left( \alpha \I + \beta \Lap \right) \mathbf{\epsilon},
\end{equation}
where $\mathbf{\epsilon}$ is the vector whose components are $\epsilon_i$. If $\mu_i$ are the eigenvalues of $\Lap$, the stability condition for the fixed point $x^*$ can be written as $\alpha + \beta \mu_i < 0$, for all $i$. Next, considering that the Laplacian matrix is semi-positive definite, the former condition in terms of the largest eigenvalue, $\mu_n$, is
\begin{equation} \label{eq:condition_L}
   \frac{1}{\mu_n} > -\left[ \slfrac{\dfrac{d g (x)}{dx}}{\dfrac{d f (x)}{dx}} \right]_{x_i = x_i^*}.
\end{equation}
Now, let us consider a second scenario for $g_j (x_{\{e_j \}})$ such that  
\begin{equation}
    g_j^* (x_{\{e_j \}}) = \frac{1}{(|e_j| - 1)} g_j (x_{\{e_j \setminus \{v_i\}\}}),
\end{equation}
which depends, for a node $v_i$, only on $x_k$, where $k \in e_j$ and $k \neq i$, and the weight is the same as for Eq.~\ref{eq:eq_separable}. Assuming that we can write $\F$ and $\G$ in the same form as Eq.~\ref{eq:cond_1}, we get 
\begin{equation}
 \dfrac{d \epsilon_i}{dt} = \alpha \epsilon_i + \sum_{e_j \in \mathcal{E}_i} \sum_{k \in e_j} \frac{\beta \epsilon_k}{|e_j| - 1},
\end{equation}
or, equivalently, 
\begin{equation}
 \dfrac{d \mathbf{\epsilon}}{dt} = \left( \alpha \I + \beta \A \right) \mathbf{\epsilon}.
\end{equation}
Denoting by $\lambda_i$ the eigenvalues of $\A$, the stability condition reads
\begin{equation} \label{eq:stability_A}
    \frac{1}{\lambda_n} < -\left[ \slfrac{\dfrac{d g (x)}{dx}}{\dfrac{d f (x)}{dx}} \right]_{x_i = x_i^*} < \frac{1}{\lambda_1}.
\end{equation}

In this two general cases, the projected adjacency matrix, Eq.~\ref{eq:adjacency}, arises naturally from the linear stability analysis around the fixed point, which means that only the projected structure is relevant for the dynamics. Moreover, as it can be seen from Eqs.~\ref{eq:condition_L} and~\ref{eq:stability_A}, we can decouple the structural and dynamical contributions to the stability conditions. Note that although the former derivations are specific for these two cases, Eq.~(\ref{eq:general_condition}) holds for a general dynamics in an arbitrary hypergraph. The framework outlined up to now is general and can be readily applied to many dynamical processes in which one is forced to go beyond pairwise interactions. In what follows, we discuss two of such applications, namely, social contagion~\cite{Iacopo2019, Arruda2019} and diffusion processes. 

\textit{Social contagion.} In this dynamical process, the state $x_i$ represents the probability of an individual to be active. The deactivation mechanism is given by $f_i(x_i) = - \delta x_i$ and the interaction functions $g_j(x_{\{e_j \}})$ are given by the product of the probability that $v_i$ is inactive and the other nodes in the hyperedge $e_j$ are active times a contact rate, $\beta_{|e_j|}$. Hence,
\begin{equation}
    g_j(x_{\{e_j \}}) =  \frac{\beta_{|e_j|}}{|e_j| - 1} (1 - x_i) \prod_{\substack{k \in e_j \\ k\neq i}} x_k.
\end{equation}
Separating the pairwise and higher-order contributions to the state of the node, we have
\begin{eqnarray} \label{eq:social_comp}
    \dfrac{d x_i}{dt} &=& -\delta x_i + \beta_{2} \sum_{\substack{e_j \in \mathcal{E}_i \\ |e_j| =2 \\ k \in e_j, k \neq i}} (1 - x_i) x_k  + \nonumber \\
    &+& \sum_{\substack{e_j \in \mathcal{E}_i \\ |e_j| > 2}} \frac{\beta_{|e_j|}}{|e_j| - 1} (1 - x_i) \prod_{\substack{k \in e_j \\ k\neq i}} x_k.
\end{eqnarray}
Assuming a symmetric fixed point, i.e., $x_i = x^*$,  
\begin{eqnarray} \label{eq:social_cont}
    \dfrac{d \epsilon_i}{dt} &=& \left( -\delta - \sum_{e_j \in \mathcal{E}_i} \frac{\beta_{|e_j|} (x^*)^{(|e_j| - 1)}}{|e_j| - 1} \right) \epsilon_i \nonumber \\
    &+& \sum_{e_j \in \mathcal{E}_i} \sum_{\substack{k \in e_j \\ k\neq i}}  \frac{\beta_{|e_j|} (1-x^*)(x^*)^{(|e_j| - 2)}}{|e_j| - 1} \epsilon_k,
\end{eqnarray}
where we can apply Eq.~\ref{eq:general_condition} to evaluate the stability of the system's dynamics. The mean-field form of Eq.~\ref{eq:social_comp} is given by
\begin{equation} \label{eq:poly}
    \dfrac{d x}{dt} = -\delta x + P^\nu(x) = -\delta x + \sum_{k=1}^\nu \beta_k c(k) (1 - x) x^{k-1},
\end{equation}
where $\nu = \max_j \{| e_j| \}$ is the maximum cardinality and $c(k)$ is the ratio between the number of hyperedges with cardinality $k$ and the number of pairwise interactions, which characterizes the structure of the hypergraph. In the steady-state this is a polynomial equation whose solutions are the fixed points of the process and their stability can be evaluated using Eq.~\ref{eq:social_cont}.

\begin{figure*}[!bt]
\includegraphics[width=0.9\linewidth]{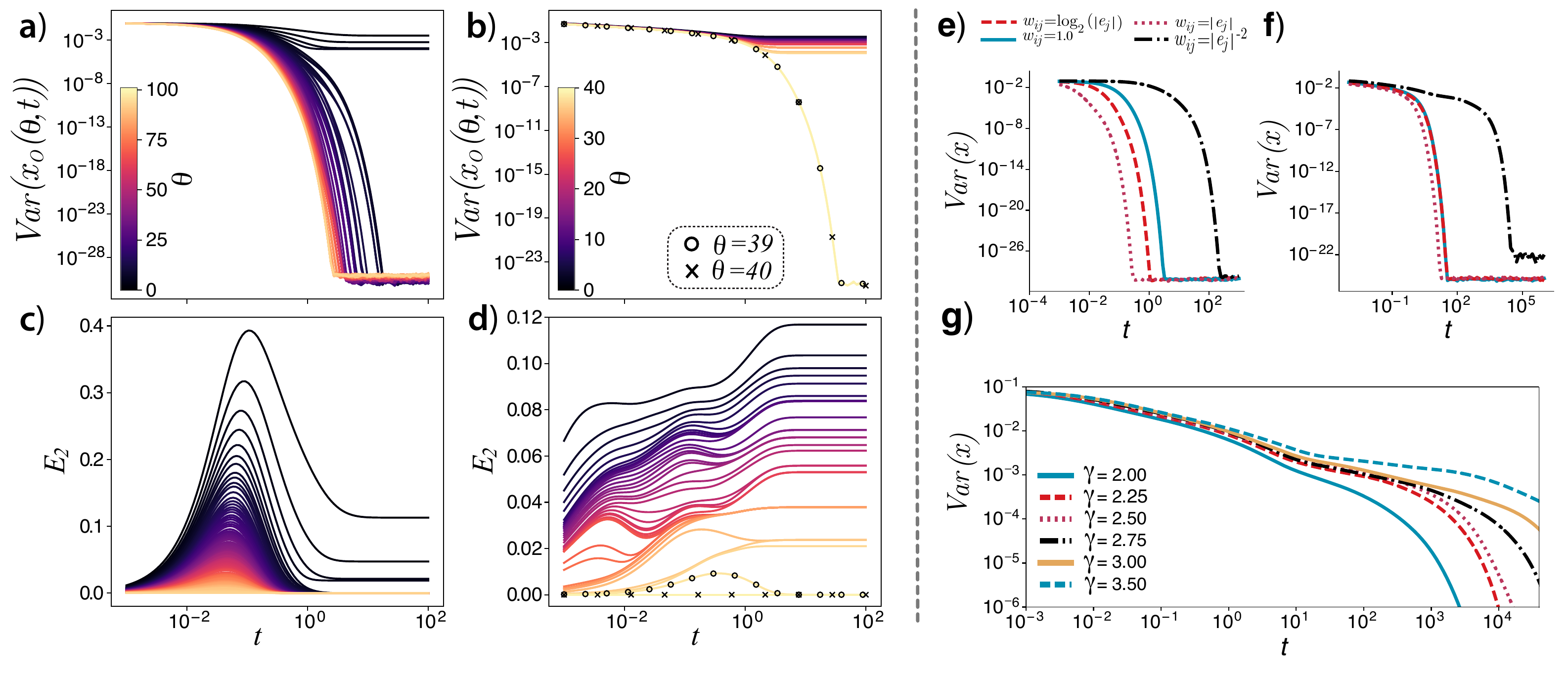}
\caption{Diffusion on real and synthetic hypergraphs. Panels (a) and (b) show the variance of states in the diffusion, while panels (c) and (d) display the error $E_2$, both as a function of time and structures considered, $\theta$ (color-coded). To represent the hypergraphs, we used in panels (a) and (c) a power-law distribution of cardinalities, whereas in (b) and (d), the hypergraph is given by the real email data (see text). The three panels to the right (e, f, and g) show results obtained for weighted diffusion on hypergraphs. Panels (e) and (f) show the effect of the different weights as a function of the cardinality, Eq.~\ref{eq:w_adjacency_m}, see the legend. Furthermore, panel (e) corresponds to a synthetic hypergraph with a power-law distribution of cardinalities, $P(|e|) \sim | e|^{-2.05}$, and (f) to the email hypergraph. Note that panel (g) explores the dependency with the exponent of the power-law weighting function.}
\label{fig:diffusion}
\end{figure*}

A fundamental observation from Eq.~\ref{eq:social_cont} is that only pairwise interactions, given by the off-diagonal terms, are responsible for the stability of the fixed point $x_i = 0$. Interestingly enough, in this scenario, the dynamics behaves like an SIS process on complex networks, and is characterized by a continuous phase transition~\cite{VanMieghem2012}. Therefore, we claim that in a social contagion model, pairwise interactions are a necessary condition for a continuous phase transition. We can support this claim using an argument by contradiction as follows. Near the absorbing state the probability for a node to be active is very small, hence, assuming an arbitrary small number $\eta \approx 0$, we can consider $x_i \in O(\eta)$. Therefore, from Eq.~\ref{eq:social_comp}, the coefficients $\beta_{|e_j|}$ are multiplying terms of order $O(\eta^{(|e_j|-1)})$.  At the steady state, $\delta x_i \sim \beta_{|e|_j} x_i$, assuming $\delta \in O(1)$ without loss of generality. From this point, we can differentiate three cases depending on $\beta_{|e_j|}$: 
\begin{enumerate}[i]
 \item If $\beta_{|e_j|} \in O(1)$ for all $|e_j| \in \{1, 2, ..., \max(|e_j|)\}$, then the leading coefficient is $\beta_2$ therefore $\beta_{2} x_i \in O(\eta)$ as assumed. In this case we have a continuous phase transition with respect to the control parameter $\beta_2$; 
 \item If $\beta_2 \in O(1)$ and $\beta_{|e_j|} \in O(\eta^{-(|e_j|-2)})$, for $|e_j| > 2$, then the leading coefficient is  $\beta_{|e_j|}  x_i  \in O(\eta)$, thus the transition can be either continuous or discontinuous;
 \item If $\beta_{|e_j|} \in O(\eta^{-(|e_j|-1)})$ for all $|e_j| \in \{1, 2, ..., \max(|e_j|)\}$, then the leading coefficient is $\beta_{|e_j|}  x_i \in O(1)$, which contradicts the initial assumption of $x_i \in O(\eta)$. This suggests a discontinuity if $\beta_2 \in O(1)$ and at least one of the higher-order coefficients is $O(\eta^{-(|e_j|-1)})$.
\end{enumerate}

To further extend the analysis in~\cite{Iacopo2019}, in order to compute $c(k)$, we consider higher-order contact patterns from two real datasets. The data were collected with proximity sensors in a high school~\cite{Mastrandrea-2015-contact, benson2018simplicial}, and a primary school~\cite{Stehl-2011-contact, benson2018simplicial}. Our results are summarized  in Fig.~\ref{fig:social_contagion} (a) and (b), where we consider the feasible fixed points and their stability. In this case, we assume $\beta_2 =1$ and $\beta_k = e^k$, in order to emphasize the role of lower and higher-order structures. Notice that, given the same parameters, we can have either continuous or discontinuous phase transitions with hysteresis, depending only on the structure of the hypergraph. Considering the graph case, i.e., $\max_j \{| e_j| \} = 2$, the social contagion model is an SIS epidemic spreading. Thus, the behavior near the fixed point $x^* = 0$ is
\begin{equation}
    \dfrac{d \epsilon_i}{dt} = -\delta \epsilon_i + \beta_{2} \sum_{e_j \in \mathcal{E}_i}  \epsilon_k.
\end{equation}
Note that the stability of the absorbing state is given by $\lambda_{\max} < \frac{\delta}{\beta_2}$, as for the quenched mean-field~\cite{Mieghem09}.

Next, we analyze the simplicial complex case, where the structure is homogeneous and consists only of pairwise interactions and triangles, hence $n=3$. In this case, Eq.~\ref{eq:poly} has the three solutions, ${x^*}_1 = 0$, and
\begin{equation} \label{eq:social_cont_sol}
    {x^*}_2^\pm = \frac{(\frac{\beta_3}{2} - \beta_2) \pm \sqrt{(\beta_2 - \frac{\beta_3}{2})^2 - 2 \beta_3 (\delta - \beta_2) }}{\beta_3}.
\end{equation}

From Eq.~\ref{eq:social_cont} we derive the following condition
\begin{equation} \label{eq:social_cont_criteria}
    \left( -\delta - \beta_2 {x^*} - \frac{\beta_3 {x^*}^2}{2} + \beta_2 (1- {x^*}) + \beta_3 (1 - {x^*}) {x^*}  \right) < 0,
\end{equation}
thus, ${x^*}_1 = 0$ is stable for $\beta_2 < \delta$. The feasibility of the solution ${x^*}_2^\pm$ was shown in~\cite{Iacopo2019}. Complementarily, Fig.~\ref{fig:social_contagion} (c) -- (f) show the results of the corresponding stability analysis, illustrating how our approach could provide new insights. As noted before, we stress that Eq.~\ref{eq:social_cont} is not limited to the case of simplicial complexes, but it can be used to further explore higher-order structures.

\textit{Diffusion process.}
Another immediate application is the diffusion process~\cite{Newman010:book}. In this case, $g_j(x_{\{e_j \}})$ is separable and $g(x_i) = x_i$. The dynamics of the process is given by
\begin{equation} \label{eq:diffusion}
 \dfrac{d \mathbf{x}}{dt} = -D \Lap \mathbf{x}.
\end{equation}
where $D$ is the diffusion constant and the minus sign is a consequence of the stability condition in Eq.~\ref{eq:condition_L}. We can arbitrarily decompose the Laplacian in $m$-uniform hypergraphs as done for the adjacency matrix in Eq.~\ref{eq:adjacency_m}. In particular, we want to consider if the lower-order contributions are enough to describe the dynamics. To this end, we can separate the Laplacian in two components $\Lap = \Lap_O(\theta) + \Lap_R(\theta)$ given by 
\begin{equation} \label{eq:Laplacian_dec}
    \Lap_O(\theta) = \sum_{m=1}^{\theta} \Lap^m  \hspace{4mm} \text{and} \hspace{4mm} \Lap_R(\theta) = \sum_{m=\theta+1}^{\max(|e_j|)} \Lap^m,
\end{equation}
where  $\theta$ differentiates between higher and lower order Laplacian, $\Lap_R(\theta)$ and $\Lap_O(\theta)$, respectively. In order to underline the important role played by the structure on the dynamics, we next analyze a diffusion process on both a synthetic hypergraph generated according to a power-law cardinality distribution, $P(|e|) \sim | e|^{-2.05}$, and a real hypergraph using data from email exchange at a European research institution~\cite{Leskovec-2007-evolution, Yin-2017-local, benson2018simplicial}. In Fig~\ref{fig:diffusion} (a) and (b) we show the variance of $x_O(\theta, t) = e^{-D\Lap_O(\theta) t}x(0)$, which is the solution of the diffusion process on the lower-order Laplacian $\Lap_O$. For a small enough $\theta$, we have disconnected components and the process does not reach the final state of the complete Laplacian. On the other hand, when the final state is reached, different $\theta$'s correspond to different time-scales, and in particular, convergence is faster when increasing $\theta$. In fact, for the synthetic case we need less information (structures) to represent the whole process, in the email dataset the convergence is only reached for $\theta-1$, as shown in Fig~\ref{fig:diffusion} (b). The relative error can be estimated as
\begin{equation} \label{eq:bound}
     E_2 = \frac{\vertii{x(t) - x_O(\theta, t)}_2}{\vertii{x(t)}_2}.
\end{equation}
For the synthetic case, as shown in Fig~\ref{fig:diffusion} (c), the error curves show peaks that are higher for lower values of $\theta$. For long eneough times, different plateau are reached depending on how many components are considered. The same is true for the email case, Fig~\ref{fig:diffusion} (d), however we have a multiple peaked pattern, suggesting that we have various structures, each one with a different time scale. As expected, only for $\theta = 39$ the error goes to zero, highlighting that real structures might have non-trivial behavior.

Another important aspect is the role on the dynamics of considering different weights for the hyperedges~\cite{Iacopo2019, Arruda2019, Unai2020}. In particular, by redefining the Laplacian with the weighted adjacency matrix, Eq.~\ref{eq:w_adjacency_m}, different choices of $w(|e_j|)$ can be used, as shown in Fig.~\ref{fig:diffusion} (e) -- (g). As expected, increasing the weights reduces the convergence time. Aside from the different time-scales in the two examples, we show that real data might also have a qualitatively different behavior, as shown in Fig.~\ref{fig:diffusion} (f) and (g). In particular, we focus on the power-law weighting functions, $w(|e_j|) = |e_j|^{-\gamma}$, in (g). In this case, since higher-order structures are heavily penalized, we observe a greater time modulation by increasing $\gamma$.

\textit{Conclusions.} 
In this paper, we have developed a framework that allows to describe the relationship between the structural organization of hypergraphs and general dynamical processes for which including more than pairwise interactions could be relevant. Within this framework, the graph projection appears naturally from a linear stability analysis around a given fixed point. Therefore, the dimensionality of the hypergraph, $N^{\max(|e_j|)}$ in its tensorial representation, is reduced to a $N^2$ space. Moreover, we showed that using a structural decomposition in $m$-uniform hypergraphs, given by Eq.~\ref{eq:adjacency_m}, enables the individual analysis of each $m$-th order component. Altogether, the methodology provides the stability condition for any general dynamical process and also allows to identify the structural components that play a role in the stability of any known fixed point. Of particular interest are the special cases of separable functions and symmetric fixed points, as one is able to separate the structural and dynamical contributions to stability. We also applied the proposed framework to two relevant dynamical processes, namely, social contagion and diffusion dynamics. For the former, we not only recovered known results for graph-based and simplicial complexes, but we obtained a necessary condition for the stability of the absorbing state, showing that that only pairwise interactions play a role for such a state. Additionally, our analysis revealed the conditions for the continuity of the phase transition also for real systems, where different kinds of bifurcations might appear. Finally, in regard to diffusion processes, we characterized the role played by the structural decomposition of the hypergraph and different weighting functions for the hyperedges on the dynamics of the system. In concluding, we stress that the framework developed here is important beyond the interest in dynamical processes, e.g., potential applications can be found in optimization problems. We are confident that this investigation will motivate further research on the spectral properties of hypergraphs, their structural characterization and in providing relevant insights in different areas where higher-order interactions can not be neglected, most notably, in brain dynamics.

\acknowledgments
We acknowledge support from Intesa Sanpaolo Innovation Center. Y. M. acknowledges partial support from the Government of Arag\'on and FEDER funds, Spain through grant ER36-20R to FENOL, and by MINECO and FEDER funds (grant FIS2017-87519-P). The funders had no role in study design, data collection, and analysis, decision to publish, or preparation of the manuscript.



\begin{thebibliography}{34}
\expandafter\ifx\csname natexlab\endcsname\relax\def\natexlab#1{#1}\fi
\expandafter\ifx\csname bibnamefont\endcsname\relax
  \def\bibnamefont#1{#1}\fi
\expandafter\ifx\csname bibfnamefont\endcsname\relax
  \def\bibfnamefont#1{#1}\fi
\expandafter\ifx\csname citenamefont\endcsname\relax
  \def\citenamefont#1{#1}\fi
\expandafter\ifx\csname url\endcsname\relax
  \def\url#1{\texttt{#1}}\fi
\expandafter\ifx\csname urlprefix\endcsname\relax\def\urlprefix{URL }\fi
\providecommand{\bibinfo}[2]{#2}
\providecommand{\eprint}[2][]{\url{#2}}

\bibitem[{\citenamefont{Boccaletti et~al.}(2006)\citenamefont{Boccaletti,
  Latora, Moreno, Chavez, and Hwang}}]{Boccaletti2008}
\bibinfo{author}{\bibfnamefont{S.}~\bibnamefont{Boccaletti}},
  \bibinfo{author}{\bibfnamefont{V.}~\bibnamefont{Latora}},
  \bibinfo{author}{\bibfnamefont{Y.}~\bibnamefont{Moreno}},
  \bibinfo{author}{\bibfnamefont{M.}~\bibnamefont{Chavez}}, \bibnamefont{and}
  \bibinfo{author}{\bibfnamefont{D.-U.} \bibnamefont{Hwang}},
  \bibinfo{journal}{Physics Reports} \textbf{\bibinfo{volume}{424}},
  \bibinfo{pages}{175 } (\bibinfo{year}{2006}), ISSN \bibinfo{issn}{0370-1573}.

\bibitem[{\citenamefont{Newman}(2010)}]{Newman010:book}
\bibinfo{author}{\bibfnamefont{M.}~\bibnamefont{Newman}},
  \emph{\bibinfo{title}{{Networks: an introduction}}}
  (\bibinfo{publisher}{Oxford University Press, Inc.}, \bibinfo{year}{2010}).

\bibitem[{\citenamefont{Pastor-Satorras
  et~al.}(2015)\citenamefont{Pastor-Satorras, Castellano, Van~Mieghem, and
  Vespignani}}]{Satorras2015}
\bibinfo{author}{\bibfnamefont{R.}~\bibnamefont{Pastor-Satorras}},
  \bibinfo{author}{\bibfnamefont{C.}~\bibnamefont{Castellano}},
  \bibinfo{author}{\bibfnamefont{P.}~\bibnamefont{Van~Mieghem}},
  \bibnamefont{and}
  \bibinfo{author}{\bibfnamefont{A.}~\bibnamefont{Vespignani}},
  \bibinfo{journal}{Rev. Mod. Phys.} \textbf{\bibinfo{volume}{87}},
  \bibinfo{pages}{925} (\bibinfo{year}{2015}).

\bibitem[{\citenamefont{Cimini et~al.}(2019)\citenamefont{Cimini, Squartini,
  Saracco, Garlaschelli, Gabrielli, and Caldarelli}}]{cimini2019}
\bibinfo{author}{\bibfnamefont{G.}~\bibnamefont{Cimini}},
  \bibinfo{author}{\bibfnamefont{T.}~\bibnamefont{Squartini}},
  \bibinfo{author}{\bibfnamefont{F.}~\bibnamefont{Saracco}},
  \bibinfo{author}{\bibfnamefont{D.}~\bibnamefont{Garlaschelli}},
  \bibinfo{author}{\bibfnamefont{A.}~\bibnamefont{Gabrielli}},
  \bibnamefont{and}
  \bibinfo{author}{\bibfnamefont{G.}~\bibnamefont{Caldarelli}},
  \bibinfo{journal}{Nature Reviews Physics} \textbf{\bibinfo{volume}{1}},
  \bibinfo{pages}{58–71} (\bibinfo{year}{2019}).

\bibitem[{\citenamefont{Lambiotte et~al.}(2019)\citenamefont{Lambiotte,
  Rosvall, and Scholtes}}]{Lambiotte2019}
\bibinfo{author}{\bibfnamefont{R.}~\bibnamefont{Lambiotte}},
  \bibinfo{author}{\bibfnamefont{M.}~\bibnamefont{Rosvall}}, \bibnamefont{and}
  \bibinfo{author}{\bibfnamefont{I.}~\bibnamefont{Scholtes}},
  \bibinfo{journal}{Nature Physics} \textbf{\bibinfo{volume}{15}},
  \bibinfo{pages}{313} (\bibinfo{year}{2019}), ISSN \bibinfo{issn}{1745-2481}.

\bibitem[{\citenamefont{Chodrow}(2019)}]{Chodrow2019}
\bibinfo{author}{\bibfnamefont{P.~S.} \bibnamefont{Chodrow}}
  (\bibinfo{year}{2019}), \eprint{arXiv:1902.09302}.

\bibitem[{\citenamefont{Petri and Barrat}(2018)}]{Petri2018}
\bibinfo{author}{\bibfnamefont{G.}~\bibnamefont{Petri}} \bibnamefont{and}
  \bibinfo{author}{\bibfnamefont{A.}~\bibnamefont{Barrat}},
  \bibinfo{journal}{Phys. Rev. Lett.} \textbf{\bibinfo{volume}{121}},
  \bibinfo{pages}{228301} (\bibinfo{year}{2018}).

\bibitem[{\citenamefont{Iacopini et~al.}(2019)\citenamefont{Iacopini, Petri,
  Barrat, and Latora}}]{Iacopo2019}
\bibinfo{author}{\bibfnamefont{I.}~\bibnamefont{Iacopini}},
  \bibinfo{author}{\bibfnamefont{G.}~\bibnamefont{Petri}},
  \bibinfo{author}{\bibfnamefont{A.}~\bibnamefont{Barrat}}, \bibnamefont{and}
  \bibinfo{author}{\bibfnamefont{V.}~\bibnamefont{Latora}},
  \bibinfo{journal}{{Nature Communications}} \textbf{\bibinfo{volume}{10}},
  \bibinfo{pages}{1} (\bibinfo{year}{2019}).

\bibitem[{\citenamefont{Armstrong}(2013)}]{armstrong2013basic}
\bibinfo{author}{\bibfnamefont{M.}~\bibnamefont{Armstrong}},
  \emph{\bibinfo{title}{Basic Topology}}, Undergraduate Texts in Mathematics
  (\bibinfo{publisher}{Springer New York}, \bibinfo{year}{2013}).

\bibitem[{\citenamefont{Estrada and Rodríguez-Velázquez}(2006)}]{Estrada2006}
\bibinfo{author}{\bibfnamefont{E.}~\bibnamefont{Estrada}} \bibnamefont{and}
  \bibinfo{author}{\bibfnamefont{J.~A.} \bibnamefont{Rodríguez-Velázquez}},
  \bibinfo{journal}{Physica A: Statistical Mechanics and its Applications}
  \textbf{\bibinfo{volume}{364}}, \bibinfo{pages}{581 } (\bibinfo{year}{2006}).

\bibitem[{\citenamefont{Ghoshal et~al.}(2009)\citenamefont{Ghoshal,
  Zlati\ifmmode~\acute{c}\else \'{c}\fi{}, Caldarelli, and
  Newman}}]{Ghoshal2009}
\bibinfo{author}{\bibfnamefont{G.}~\bibnamefont{Ghoshal}},
  \bibinfo{author}{\bibfnamefont{V.}~\bibnamefont{Zlati\ifmmode~\acute{c}\else
  \'{c}\fi{}}}, \bibinfo{author}{\bibfnamefont{G.}~\bibnamefont{Caldarelli}},
  \bibnamefont{and} \bibinfo{author}{\bibfnamefont{M.~E.~J.}
  \bibnamefont{Newman}}, \bibinfo{journal}{Phys. Rev. E}
  \textbf{\bibinfo{volume}{79}}, \bibinfo{pages}{066118}
  (\bibinfo{year}{2009}).

\bibitem[{\citenamefont{Bod{\'o} et~al.}(2016)\citenamefont{Bod{\'o}, Katona,
  and Simon}}]{Bodo2016}
\bibinfo{author}{\bibfnamefont{{\'A}.}~\bibnamefont{Bod{\'o}}},
  \bibinfo{author}{\bibfnamefont{G.~Y.} \bibnamefont{Katona}},
  \bibnamefont{and} \bibinfo{author}{\bibfnamefont{P.~L.} \bibnamefont{Simon}},
  \bibinfo{journal}{Bulletin of Mathematical Biology}
  \textbf{\bibinfo{volume}{78}}, \bibinfo{pages}{713} (\bibinfo{year}{2016}).

\bibitem[{\citenamefont{Banerjee et~al.}(2017)\citenamefont{Banerjee, Char, and
  Mondal}}]{Banerjee2017}
\bibinfo{author}{\bibfnamefont{A.}~\bibnamefont{Banerjee}},
  \bibinfo{author}{\bibfnamefont{A.}~\bibnamefont{Char}}, \bibnamefont{and}
  \bibinfo{author}{\bibfnamefont{B.}~\bibnamefont{Mondal}}
  (\bibinfo{year}{2017}), \eprint{1601.02136v4}.

\bibitem[{\citenamefont{Benson et~al.}(2018)\citenamefont{Benson, Abebe,
  Schaub, Jadbabaie, and Kleinberg}}]{benson2018simplicial}
\bibinfo{author}{\bibfnamefont{A.~R.} \bibnamefont{Benson}},
  \bibinfo{author}{\bibfnamefont{R.}~\bibnamefont{Abebe}},
  \bibinfo{author}{\bibfnamefont{M.~T.} \bibnamefont{Schaub}},
  \bibinfo{author}{\bibfnamefont{A.}~\bibnamefont{Jadbabaie}},
  \bibnamefont{and}
  \bibinfo{author}{\bibfnamefont{J.}~\bibnamefont{Kleinberg}},
  \bibinfo{journal}{Proceedings of the National Academy of Sciences}
  \textbf{\bibinfo{volume}{115}}, \bibinfo{pages}{E11221}
  (\bibinfo{year}{2018}).

\bibitem[{\citenamefont{Banerjee}(2019)}]{Banerjee2019}
\bibinfo{author}{\bibfnamefont{A.}~\bibnamefont{Banerjee}}
  (\bibinfo{year}{2019}), \eprint{1711.09356v3}.

\bibitem[{\citenamefont{Ouvrard et~al.}(2017)\citenamefont{Ouvrard, Goff, and
  Marchand-Maillet}}]{Ouvrard2018}
\bibinfo{author}{\bibfnamefont{X.}~\bibnamefont{Ouvrard}},
  \bibinfo{author}{\bibfnamefont{J.-M.~L.} \bibnamefont{Goff}},
  \bibnamefont{and}
  \bibinfo{author}{\bibfnamefont{S.}~\bibnamefont{Marchand-Maillet}}
  (\bibinfo{year}{2017}), \eprint{arXiv:1712.08189}.

\bibitem[{\citenamefont{de~Arruda et~al.}(2020)\citenamefont{de~Arruda, Petri,
  and Moreno}}]{Arruda2019}
\bibinfo{author}{\bibfnamefont{G.~F.} \bibnamefont{de~Arruda}},
  \bibinfo{author}{\bibfnamefont{G.}~\bibnamefont{Petri}}, \bibnamefont{and}
  \bibinfo{author}{\bibfnamefont{Y.}~\bibnamefont{Moreno}},
  \bibinfo{journal}{Phys. Rev. Research} \textbf{\bibinfo{volume}{2}},
  \bibinfo{pages}{023032} (\bibinfo{year}{2020}).

\bibitem[{\citenamefont{Alvarez-Rodriguez
  et~al.}(2020)\citenamefont{Alvarez-Rodriguez, Battiston, de~Arruda, Moreno,
  Perc, and Latora}}]{Unai2020}
\bibinfo{author}{\bibfnamefont{U.}~\bibnamefont{Alvarez-Rodriguez}},
  \bibinfo{author}{\bibfnamefont{F.}~\bibnamefont{Battiston}},
  \bibinfo{author}{\bibfnamefont{G.~F.} \bibnamefont{de~Arruda}},
  \bibinfo{author}{\bibfnamefont{Y.}~\bibnamefont{Moreno}},
  \bibinfo{author}{\bibfnamefont{M.}~\bibnamefont{Perc}}, \bibnamefont{and}
  \bibinfo{author}{\bibfnamefont{V.}~\bibnamefont{Latora}}
  (\bibinfo{year}{2020}), \eprint{arXiv:2001.10313}.

\bibitem[{\citenamefont{Carletti et~al.}(2020)\citenamefont{Carletti,
  Battiston, Cencetti, and Fanelli}}]{Carletti2019}
\bibinfo{author}{\bibfnamefont{T.}~\bibnamefont{Carletti}},
  \bibinfo{author}{\bibfnamefont{F.}~\bibnamefont{Battiston}},
  \bibinfo{author}{\bibfnamefont{G.}~\bibnamefont{Cencetti}}, \bibnamefont{and}
  \bibinfo{author}{\bibfnamefont{D.}~\bibnamefont{Fanelli}},
  \bibinfo{journal}{Phys. Rev. E} \textbf{\bibinfo{volume}{101}},
  \bibinfo{pages}{022308} (\bibinfo{year}{2020}).

\bibitem[{\citenamefont{Qi and Luo}(2017)}]{qi2017}
\bibinfo{author}{\bibfnamefont{L.}~\bibnamefont{Qi}} \bibnamefont{and}
  \bibinfo{author}{\bibfnamefont{Z.}~\bibnamefont{Luo}},
  \emph{\bibinfo{title}{Tensor Analysis: Spectral Theory and Special Tensors}},
  Other Titles in Applied Mathematics (\bibinfo{publisher}{Society for
  Industrial and Applied Mathematics}, \bibinfo{year}{2017}), ISBN
  \bibinfo{isbn}{9781611974744}.

\bibitem[{\citenamefont{Ouvrard}(2020)}]{ouvrard2020}
\bibinfo{author}{\bibfnamefont{X.}~\bibnamefont{Ouvrard}}
  (\bibinfo{year}{2020}), \eprint{arXiv:2002.05014}.

\bibitem[{\citenamefont{{Karypis} et~al.}(1999)\citenamefont{{Karypis},
  {Aggarwal}, {Kumar}, and {Shekhar}}}]{Karypis1999}
\bibinfo{author}{\bibfnamefont{G.}~\bibnamefont{{Karypis}}},
  \bibinfo{author}{\bibfnamefont{R.}~\bibnamefont{{Aggarwal}}},
  \bibinfo{author}{\bibfnamefont{V.}~\bibnamefont{{Kumar}}}, \bibnamefont{and}
  \bibinfo{author}{\bibfnamefont{S.}~\bibnamefont{{Shekhar}}},
  \bibinfo{journal}{IEEE Transactions on Very Large Scale Integration (VLSI)
  Systems} \textbf{\bibinfo{volume}{7}} (\bibinfo{year}{1999}), ISSN
  \bibinfo{issn}{1557-9999}.

\bibitem[{\citenamefont{Tian et~al.}(2009)\citenamefont{Tian, Hwang, and
  Kuang}}]{Tian2009}
\bibinfo{author}{\bibfnamefont{Z.}~\bibnamefont{Tian}},
  \bibinfo{author}{\bibfnamefont{T.}~\bibnamefont{Hwang}}, \bibnamefont{and}
  \bibinfo{author}{\bibfnamefont{R.}~\bibnamefont{Kuang}},
  \bibinfo{journal}{Bioinformatics} \textbf{\bibinfo{volume}{25}},
  \bibinfo{pages}{2831} (\bibinfo{year}{2009}), ISSN \bibinfo{issn}{1367-4803}.

\bibitem[{\citenamefont{Voloshin}(2013)}]{Voloshin2013}
\bibinfo{author}{\bibfnamefont{V.~I.} \bibnamefont{Voloshin}}
  (\bibinfo{year}{2013}).

\bibitem[{\citenamefont{Bretto}(2013)}]{Bretto2013}
\bibinfo{author}{\bibfnamefont{A.}~\bibnamefont{Bretto}},
  \emph{\bibinfo{title}{Hypergraph Theory: An Introduction}}
  (\bibinfo{publisher}{Springer Publishing Company, Incorporated},
  \bibinfo{year}{2013}), ISBN \bibinfo{isbn}{3319000799}.

\bibitem[{\citenamefont{{Valdivia} et~al.}(2019)\citenamefont{{Valdivia},
  {Buono}, {Plaisant}, {Dufournaud}, and {Fekete}}}]{Valdivia2019}
\bibinfo{author}{\bibfnamefont{P.}~\bibnamefont{{Valdivia}}},
  \bibinfo{author}{\bibfnamefont{P.}~\bibnamefont{{Buono}}},
  \bibinfo{author}{\bibfnamefont{C.}~\bibnamefont{{Plaisant}}},
  \bibinfo{author}{\bibfnamefont{N.}~\bibnamefont{{Dufournaud}}},
  \bibnamefont{and} \bibinfo{author}{\bibfnamefont{J.}~\bibnamefont{{Fekete}}},
  \bibinfo{journal}{IEEE Transactions on Visualization and Computer Graphics}
  pp. \bibinfo{pages}{1--1} (\bibinfo{year}{2019}), ISSN
  \bibinfo{issn}{2160-9306}.

\bibitem[{\citenamefont{Payne}(2019)}]{Payne2019}
\bibinfo{author}{\bibfnamefont{J.}~\bibnamefont{Payne}} (\bibinfo{year}{2019}),
  \eprint{arXiv:1910.02633}.

\bibitem[{\citenamefont{Jiang et~al.}(2019)\citenamefont{Jiang, Wei, Feng, Cao,
  and Gao}}]{Jiang2019}
\bibinfo{author}{\bibfnamefont{J.}~\bibnamefont{Jiang}},
  \bibinfo{author}{\bibfnamefont{Y.}~\bibnamefont{Wei}},
  \bibinfo{author}{\bibfnamefont{Y.}~\bibnamefont{Feng}},
  \bibinfo{author}{\bibfnamefont{J.}~\bibnamefont{Cao}}, \bibnamefont{and}
  \bibinfo{author}{\bibfnamefont{Y.}~\bibnamefont{Gao}}, in
  \emph{\bibinfo{booktitle}{Proceedings of the Twenty-Eighth International
  Joint Conference on Artificial Intelligence, {IJCAI-19}}}
  (\bibinfo{publisher}{International Joint Conferences on Artificial
  Intelligence Organization}, \bibinfo{year}{2019}), pp.
  \bibinfo{pages}{2635--2641}.

\bibitem[{\citenamefont{Mieghem}(2012)}]{VanMieghem2012}
\bibinfo{author}{\bibfnamefont{P.~V.} \bibnamefont{Mieghem}},
  \bibinfo{journal}{{EPL} (Europhysics Letters)} \textbf{\bibinfo{volume}{97}},
  \bibinfo{pages}{48004} (\bibinfo{year}{2012}).

\bibitem[{\citenamefont{Mastrandrea et~al.}(2015)\citenamefont{Mastrandrea,
  Fournet, and Barrat}}]{Mastrandrea-2015-contact}
\bibinfo{author}{\bibfnamefont{R.}~\bibnamefont{Mastrandrea}},
  \bibinfo{author}{\bibfnamefont{J.}~\bibnamefont{Fournet}}, \bibnamefont{and}
  \bibinfo{author}{\bibfnamefont{A.}~\bibnamefont{Barrat}},
  \bibinfo{journal}{{PLOS} {ONE}} \textbf{\bibinfo{volume}{10}},
  \bibinfo{pages}{e0136497} (\bibinfo{year}{2015}).

\bibitem[{\citenamefont{Stehl{\'{e}} et~al.}(2011)\citenamefont{Stehl{\'{e}},
  Voirin, Barrat, Cattuto, Isella, Pinton, Quaggiotto, den Broeck, R{\'{e}}gis,
  Lina et~al.}}]{Stehl-2011-contact}
\bibinfo{author}{\bibfnamefont{J.}~\bibnamefont{Stehl{\'{e}}}},
  \bibinfo{author}{\bibfnamefont{N.}~\bibnamefont{Voirin}},
  \bibinfo{author}{\bibfnamefont{A.}~\bibnamefont{Barrat}},
  \bibinfo{author}{\bibfnamefont{C.}~\bibnamefont{Cattuto}},
  \bibinfo{author}{\bibfnamefont{L.}~\bibnamefont{Isella}},
  \bibinfo{author}{\bibfnamefont{J.-F.} \bibnamefont{Pinton}},
  \bibinfo{author}{\bibfnamefont{M.}~\bibnamefont{Quaggiotto}},
  \bibinfo{author}{\bibfnamefont{W.~V.} \bibnamefont{den Broeck}},
  \bibinfo{author}{\bibfnamefont{C.}~\bibnamefont{R{\'{e}}gis}},
  \bibinfo{author}{\bibfnamefont{B.}~\bibnamefont{Lina}}, \bibnamefont{et~al.},
  \bibinfo{journal}{{PLoS} {ONE}} \textbf{\bibinfo{volume}{6}},
  \bibinfo{pages}{e23176} (\bibinfo{year}{2011}).

\bibitem[{\citenamefont{Mieghem et~al.}(2009)\citenamefont{Mieghem, Omic, and
  Kooij}}]{Mieghem09}
\bibinfo{author}{\bibfnamefont{P.~V.} \bibnamefont{Mieghem}},
  \bibinfo{author}{\bibfnamefont{J.}~\bibnamefont{Omic}}, \bibnamefont{and}
  \bibinfo{author}{\bibfnamefont{R.}~\bibnamefont{Kooij}},
  \bibinfo{journal}{IEEE/ACM Trans. Netw.} \textbf{\bibinfo{volume}{17}},
  \bibinfo{pages}{1} (\bibinfo{year}{2009}).

\bibitem[{\citenamefont{Leskovec et~al.}(2007)\citenamefont{Leskovec,
  Kleinberg, and Faloutsos}}]{Leskovec-2007-evolution}
\bibinfo{author}{\bibfnamefont{J.}~\bibnamefont{Leskovec}},
  \bibinfo{author}{\bibfnamefont{J.}~\bibnamefont{Kleinberg}},
  \bibnamefont{and}
  \bibinfo{author}{\bibfnamefont{C.}~\bibnamefont{Faloutsos}},
  \bibinfo{journal}{{ACM} Transactions on Knowledge Discovery from Data}
  \textbf{\bibinfo{volume}{1}} (\bibinfo{year}{2007}).

\bibitem[{\citenamefont{Yin et~al.}(2017)\citenamefont{Yin, Benson, Leskovec,
  and Gleich}}]{Yin-2017-local}
\bibinfo{author}{\bibfnamefont{H.}~\bibnamefont{Yin}},
  \bibinfo{author}{\bibfnamefont{A.~R.} \bibnamefont{Benson}},
  \bibinfo{author}{\bibfnamefont{J.}~\bibnamefont{Leskovec}}, \bibnamefont{and}
  \bibinfo{author}{\bibfnamefont{D.~F.} \bibnamefont{Gleich}}, in
  \emph{\bibinfo{booktitle}{Proceedings of the 23rd {ACM} {SIGKDD}
  International Conference on Knowledge Discovery and Data Mining}}
  (\bibinfo{publisher}{{ACM} Press}, \bibinfo{year}{2017}).

\end{thebibliography}

\end{document}